# Structural dependence of the molecular mobility in the amorphous fractions of polylactide

Nicolas Delpouve[1]*, Laurent Delbreilh[1], Grégory Stoclet[2], Allisson Saiter[1], Eric Dargent[1].

[1] Normandie Univ, France ; Université and INSA Rouen ; AMME-LECAP EA 4528 International Lab., Av. de l'Université, BP12, 76801 St Etienne du Rouvray, France.

[2] Université de Lille Nord de France, UMR CNRS 8207, Unité Matériaux et Transformations, Université Lille1 Sciences et Technologies, Bâtiment C6, 59655 Villeneuve d'Ascq, France.

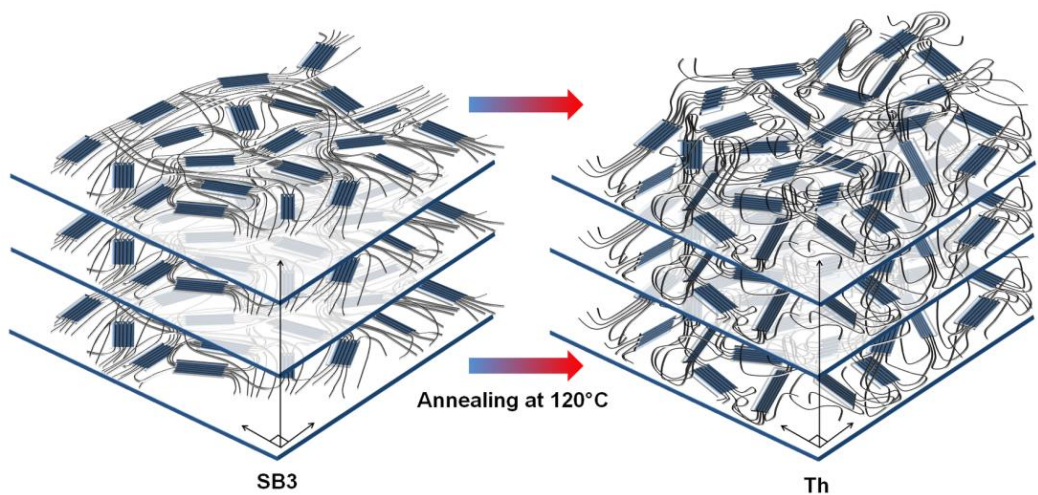

**KEYWORDS:** Drawn Polylactide, Cooperativity, Heterogeneity, CRR, Fragility, Amorphous phase, Crystallization.

AUTHOR INFORMATION

Corresponding author: Tel: 0033 2 32 95 51 65; fax: 0033 2 32 95 50 82; nicolas.delpouve1@univ-rouen.fr

## ABSTRACT

Fragility index and cooperativity length characterizing the molecular mobility in the amorphous phase are for the first time calculated in drawn polylactide (PLA). The microstructure of the samples is investigated from wide-angle X-ray scattering (WAXS) whereas the amorphous phase dynamics are revealed from broadband dielectric spectroscopy (BDS) and temperature-modulated differential scanning calorimetry (TMDSC). The drawing processes induce the decrease of both cooperativity and fragility with the orientation of the macromolecules. Post-drawing annealing reveals an unusual absence of correlation between the evolutions of cooperativity length and fragility. The cooperativity length remains the same compared to the drawn sample while a huge increase of the fragility index is recorded. By splitting the fragility index in a volume contribution and an energetic contribution, it is revealed that the amorphous phase in annealed samples exhibits a high energetic parameter, even exceeding the amorphous matrix value. It is assumed that the relaxation process is driven in such a way that the volume hindrance caused by the thermomechanical constraint is compensated by the acceleration of segmental motions linked to the increase of degrees of freedom. This result should also contribute to the understanding of the constraint slackening in the amorphous phase during annealing of drawn PLA, which causes among others the decrease of its barrier properties.

## INTRODUCTION

As one of the most promising biopolymers, polylactide (PLA) has exhibited promising applicability in the past decades due to its performance in renewability, biodegradability, biocompatibility, and mechanical properties.[1,2] Its microstructure has been widely studied and it has been shown that PLA can be obtained in many different grades as a function of the stereoregularity,[3] including stereocomplexes.[4]
Moreover, PLA exhibits at least four different crystalline forms ($\alpha$, $\beta$, $\gamma$, and $\delta$) depending on the crystallization conditions.[5,6,7,8,9] Texturing via plastic deformation is well-known to improve stiffness, strength and barrier properties of semicrystalline polymers. The structural characterization via *ex situ* X-ray measurements showed that strain-induced crystallization into the so-called disordered $\alpha$-form (i.e., $\alpha$' form—now called the $\delta$ form)[9] occurred upon drawing at 90 °C (i.e., about 30 °C above the glass transition temperature $T_g$). Otherwise, drawing at $T_g$ + 10 °C results in a structural ordering of the mesomorphic type. Between these two temperatures, both crystal and mesophase were shown to coexist.[10,11] Moreover, semicrystalline PLA is often described following a three-phase model that is containing the crystalline phase, the mobile amorphous phase (MAP) and the rigid amorphous fraction (RAF).[12,13,14] The RAF is the result of the strong restrictions of the amorphous chain segment mobility, due to the fixation of the polymer chain to the crystalline lamella.[15] Unlike the MAP, the RAF does not relax at the glass transition but devitrifies in a temperature domain situated between the glass transition and the fusion. In the case of PLA, the devitrification domain of the RAF is close to its formation temperature (i.e., the temperature of crystallization).[16]

The macromolecule arrangement generally plays a key role in the transport properties, and the small molecules diffusing actually behave as molecular probes. Unusually in the case of PLA, contradictory results are found in the literature concerning the variations of barrier properties with crystallinity. As an example, Drieskens et al. found that crystallization of PLA caused a decrease of the $O_2$ permeability, but not in linear proportion with the decrease in amorphous volume.[17] On the other hand, Colomines et al. reported that crystallinity had no effect on the $O_2$ barrier property of PLA.[18] Some authors have suggested that the barrier properties are dependent on the polymorphic structure of PLA due to differences of molecular packing between $\delta$ and $\alpha$ crystalline forms.[19] This result is supported by the PALS observations of Del Rio et al. in which a dedensification of the amorphous phase is identified mainly during $\delta$ crystallization due to RAF creation.[20] In a previous work involving different drawing treatments we have pointed out that the organization of both crystalline and amorphous phases plays a more important role than the degree of crystallinity in controlling the water barrier properties.[21] Our conclusions were recently confirmed by the work of Bai et al. who succeeded in designing a parallel-aligned shish-kebab-like microstructure with well-interlocked boundaries to get a strong improvement of PLA gas barrier properties.[22] All these studies reveal the complexity of the amorphous phase organization related to the microstructure changes and the strong urge to investigate this inter-dependence in more detail.

The molecular mobility in the amorphous phase is mainly driven by intermolecular interactions and cooperativity between chains. To investigate these parameters, some models have been developed based on the study of the α relaxation frequency dependence[23] and its dissipation zone broadness in frequency[24] and in temperature.[25,26,27] The steepness of temperature dependence of the relaxation rate at $T_g$ is associated in literature to fragility index, $m$, as defined by Angell et al.[28] As it characterizes the vitrification properties of a glass-forming system, fragility is considered as "a key parameter for observing modifications of the relaxation environment of the macromolecules". [29] The intermolecular interactions play, for most of polymeric systems, a major role in the viscous slowing down of molecular dynamics when cooling down a liquid glass-former close to $T_g$. Several models have been developed to study these coupling interactions by defining parameters, such as configurational entropy,[30,31] coupling parameter[32] or cooperativity length scale of molecular motions.[33] The evolution of the cooperativity length in glass-forming liquids with time and temperature always attracts the interest of researchers.[34,35] The idea of a growing characteristic length scale influencing relaxation dynamics of glass-forming liquids[36,37] has been introduced with the concept of cooperative rearranging region (CRR) defined as the smallest amorphous domain where a conformational rearrangement may occur without causing rearrangements in the surrounding.[33] In recent years, the evolution of the cooperativity length upon varying external parameters as well as molecular characteristics has been experimentally studied to understand how cooperativity length correlates with other relaxation parameters—such as fragility, glass transition temperature, and relaxation time. In this work, we propose to investigate the nature of the amorphous phase in semicrystalline PLA having undergone different thermomechanical treatments, by calculating both the fragility of the glass-forming liquid and the cooperativity length at the dynamic glass transition. This work aims in particular at understanding for PLA the link between molecular mobility in the amorphous phase and its microstructure. Several microstructures have been investigated using different path of crystallization i.e. thermal crystallization, different strain induced crystallization processes and combination of both thermal and strain-induced processes. To minimize effects related to the polymorphism of the crystalline phase, thermal crystallization and drawing have been carried in the same temperature domain. The discussion provides new insights concerning the complex correlation between the microstructure and the molecular dynamics of the amorphous phase.

## EXPERIMENTAL SECTION

### Materials

Polylactide (PLA) pellets were provided by Natureworks® (grade 4042D). The content of L-lactide was about 95.7%. The number-average and weight-average molecular weights were Mn = 116 kDa and Mw = 188 kDa, respectively, as measured by gel permeation chromatography. Pellets were compression-molded for 15 minutes in a SCAMEX® 20 T into films between steel plates at 185 °C under a pressure of 100 bars, before being quenched to 0°C in order to avoid crystallization. The amorphous character ($A$) of the PLA film was controlled using differential scanning calorimetry (DSC) and wide-angle X-ray scattering (WAXS). The glass transition temperature and melting temperature, determined by means of DSC, were located around 60 °C and 155 °C, respectively. Crystallization procedures were carried on amorphous films. One sample was annealed at 80°C (*TCR*); the others were drawn.

### Drawing Processes

Drawing experiments were carried out on a Cellier tenter frame consisting of four pantographs each equipped with ten pneumatic grips. The two movable pantographs were driven by hydraulic jacks. Square specimens of 100 x 100 mm² gauge widths were used. Samples were drawn at $T_d$ = 70 °C (i.e., between the glass transition and the cold crystallization temperatures), at a constant jack speed of 1 mm/s (i.e., an initial stretching speed of 0.01 s⁻¹). The true strains of the samples were determined from a square grid of 10 x 10 mm² mesh size printed on the sample prior to drawing.

Three drawing modes were used in this study. In the Uniaxial Constant Width (*UCW*) drawing mode, the films are drawn only in one direction (called the machine direction [MD]) while the film is constrained in the perpendicular direction (called the transverse direction [TD]). In the Simultaneous Biaxial (*SB*) drawing mode, the films are simultaneously drawn in two perpendicular directions at the same rate and with the same ratio. During the Sequential Biaxial drawing (*SEQ*), the films are first drawn in MD in *UCW* mode. Then the films are drawn in TD in *UCW* mode. The same stretching rate is used in MD and TD. The draw ratio λ in one direction is defined as λ = L/$L_0$, where $L_0$ is the gauge length and L is the macroscopic sample length assessed from the jack displacement.

The code for each drawn sample describes the draw ratios achieved in each direction (i.e., $\lambda_{MD}$x$\lambda_{TD}$). In this study, the samples studied are, respectively, *3x1 (UCW); 3x3 (SEQ); and 2x2, 3x3, 4x4 (SB)*. A last sample, previously drawn in SB drawing mode with a draw ratio of 3x3, was then thermofixed (*Th*) (i.e., annealed in an oven at 120 °C [$T_g$ + 60 °C] for ten minutes).

**Wide-Angle X-Ray Scattering (WAXS)**

WAXS analyses were carried out owing to a PANalytical sealed tube operating at 40kV and 20mA. The Cu-Kα radiation (λ = 1.54Å) was selected with a nickel filter. The WAXS patterns were recorded on a 2-D CCD camera (Photonic Science). Through-views were collected during simultaneous and sequential drawings. In order to determine the strain-induced structure, a semiquantitative analysis was performed starting from the 180°-azimuthally integrated profiles calculated using the FIT2D® software. The WAXS intensity profiles were further deconvoluted using the PeakFit® software, assuming Gaussian profiles for all scattering peaks and amorphous halos. The weight fraction of the various phases was computed from the ratio of the specific scattering contribution of each to the total scattering area as described elsewhere.[10] The amorphous intensity profile in semicrystalline samples was identified to the one of the thoroughly amorphous undrawn sample, regarding both the position and the full width at half maximum (FWHM).

**Temperature-Modulated Differential Scanning Calorimetry (TMDSC)**

Experiments were performed on a Thermal Analysis® instrument (TA DSC 2920). Nitrogen was used as purge gas (70 mL/min). The sample weights were about 2 mg, encapsulated in standard DSC aluminum alloy pans, and disposed in a way to have the best thermal contact possible. Before the experiments, samples were stored in vacuum desiccators over $P_2O_5$ for at least two weeks to avoid moisture-sorption effects. Calibration in temperature was carried out using standard values of indium and benzophenone. Calibration in energy was carried out using standard values of indium, and the specific heat capacity for each sample was measured using sapphire as a reference. We chose heat-iso modulation parameters (oscillation amplitude of 0.318 K, oscillation period of 60 s, and heating rate of 2 K.min$^{-1}$) as recommended for simultaneously studying glass transition, cold crystallization and fusion in semicrystalline polymers.[38] The complete deconvolution procedure proposed by Reading et al.[38] was applied in these experiments. The in-phase component C' and out-of-phase component C" of the complex heat capacity were then obtained.

**Broadband Dielectric Spectroscopy (BDS)**

Dielectric relaxation spectra were measured with an Alpha Analyzer from Novocontrol (frequency interval: $10^{-2}$—$10^6$ Hz). A film of the studied material was placed between parallel electrodes, and the temperature was controlled through a heated flow of nitrogen gas, by means of a Quatro Cryosystem, from -160 °C to 160 °C. During the whole period of the measurement, the sample was kept in a pure nitrogen atmosphere. To analyze dielectric relaxation curves, Havriliak-Negami (HN) complex function was used.[39]

$$\varepsilon^*(\omega) = \varepsilon_\infty + \frac{\Delta\varepsilon_{HN}}{\left[1 + (i\omega\tau_{HN})^{\alpha_{HN}}\right]^{\beta_{HN}}} \quad (1)$$

This formalism allows fitting the real (ε' [ω]) and imaginary (ε" [ω]) parts of the complex dielectric permittivity data (ε*[ω]) with the following equations:

$$\varepsilon'(\omega) = \varepsilon_\infty + \Delta\varepsilon_{HN} \frac{\cos(\beta_{HN}\varphi_{HN})}{\left(1 + 2\sin\left(\frac{\pi(1-\alpha_{HN})}{2}\right)(\omega\tau_{HN})^{\alpha_{HN}} + (\omega\tau_{HN})^{2\alpha_{HN}}\right)^{\beta_{HN}/2}} \quad (2)$$

$$\varepsilon''(\omega) = \Delta\varepsilon_{HN} \frac{\sin(\beta_{HN}\varphi_{HN})}{\left(1 + 2\sin\left(\frac{\pi(1-\alpha_{HN})}{2}\right)(\omega\tau_{HN})^{\alpha_{HN}} + (\omega\tau_{HN})^{2\alpha_{HN}}\right)^{\beta_{HN}/2}} \quad (3)$$

With
$$\varphi_{HN} = \arctan\left(\frac{(\omega\tau_{HN})^{\alpha_{HN}}\cos\left(\frac{\pi(1-\alpha_{HN})}{2}\right)}{1+(\omega\tau_{HN})^{\alpha_{HN}}\sin\left(\frac{\pi(1-\alpha_{HN})}{2}\right)}\right)$$

and $\omega = 2\pi f$ (4)

Where f is the frequency, $\omega$ is the angular pulsation, $\Delta\varepsilon_{HN}$ is the relaxation strength, $\tau_{HN}$ is the relaxation time, and $\alpha_{HN}$ and $\beta_{HN}$ are the symmetric and asymmetric broadening factors.

Conduction effects were treated in the usual way by adding a contribution $\sigma''_{cond} = \sigma_0/[\omega^s\varepsilon_0]$ to the dielectric loss, where $\sigma_0$ is related to the specific direct current (dc) conductivity of the sample. The parameter s ($0 < s \leq 1$) describes s = 1 Ohmic and s < 1 non-Ohmic effects in the conductivity. If two relaxation processes were observed in the experimental frequency window, a sum of two HN functions was fitted to the experimental data. The fitting procedure has been conducted on the loss part of the signal and also on the real part (without conductivity contribution) in order to improve the accuracy of the resulting fitting parameters.

## RESULTS AND DISCUSSION

### Microstructural Characterization

Ex-situ WAXS patterns obtained for sequential or simultaneous biaxial drawings are presented in Figure 1 and placed on the stress/draw ratio curves. For the initial undrawn material (sample *A*), we observe a diffuse amorphous halo characteristic of a fully isotropic amorphous material. During *SB* drawing (Figure 1a), the plastic deformation first proceeds at constant stress until the draw ratio $\lambda$ = 2.5 x 2.5. Then a strain-hardening phenomenon is observed until sample break around $\lambda$ = 4x4. Hence, regarding the *SB*-drawn material with a draw ratio of 2 x 2 (*SB2*), the WAXS pattern exhibits a weak diffraction ring having a low intensity, indicating, first, that a small amount of crystals has been induced upon drawing and, second, that the strain-induced crystalline structure is nearly isotropic. For higher draw ratios, the increase of stress with deformation is correlated to a strain-induced crystallization phenomenon. The *SB3* pattern shows a clear intensification of the main diffraction ring associated with the appearance of a second external ring, meaning that the degree of crystallinity increases with the draw ratio and that crystalline distribution remains nearly isotropic in the plane film. On the contrary, in the case of sequential drawing (Figure 1b), assimilated to successive *UCW* drawings in MD and TD, respectively, preferential crystalline orientations are detected. After the first drawing stage (*UCW* in MD), the strain hardening occurring above $\lambda$ = 2.5 x 1 is related to a strain-induced crystallization phenomenon.[10] Worth noting is that the strain-induced crystalline structure is strongly oriented as revealed by the two intense equatorial diffraction arcs that indicate crystals are oriented along the draw direction. During the second stage, the hardening continues until $\lambda$ = 3 x 2.5; then the deformation occurs at constant stress. When TD drawing is over (*SEQ*), the diffraction arcs are weaker and a diffraction ring appears on the WAXS pattern. Nevertheless, four intensifications are located along the meridian and equatorial directions, corresponding to the two drawing directions. We note that these microstructure differences do not affect much mechanical properties as stress values are similar for the *SB3* and *SEQ* materials. The integrated intensity profiles for these two materials have been computed from the WAXS patterns, and the *SB3* resulting spectrum is reported in Figure 2. Two peaks appear at $2\theta = 16.4°$ and $2\theta = 18.6°$. These two peaks are respectively ascribed to the (200)/100 and (203) diffraction planes of the α′ crystalline form[7,8] (i.e., the disordered α form) —now called the δ form.[9] Thus the crystalline phase induced upon the biaxial drawing is the same as the one obtained in the case of uniaxial drawing in equivalent draw conditions.[10,11] This is in good agreement with previous results, showing that this crystalline form is induced for samples thermally or mechanically crystallized at temperatures below 120 °C.[11] The deconvolution of the integrated intensity profiles shows that the addition of a mesomorphic extra contribution, previously observed at $2\theta = 16.2°$ in uniaxially drawn samples,[10] is also required to fit the experimental data for biaxially drawn materials. The mesophase is an intermediate ordering fraction which is clearly identifiable from WAXS analyses.[10,40,41] Its FWHM value, which is equal to 3.5° evidences an intermediate ordering between the crystalline state which FWHM is about 0.5° and the amorphous state which fwhm is about 8°.[11] The average interchain distance in the mesomorphic form is substantially reduced with regard to the most probable spacing of the amorphous chains.[10] The weight fraction of each phase has been calculated from the intensity profiles and from the average heat flow curves (shown in a previous paper[21]) obtained by means of TMDSC. Both techniques lead to similar results; TMDSC results are presented in Table 1.

The understanding of the correlation between mesophase and Rigid Amorphous Fraction (RAF) is still a complex subject to debate.[42,43] Most often, the RAF is identified from calorimetric studies whereas the mesophase is quantified from WAXS analyses. To date, it is however not possible by means of these techniques to separate the rigid amorphous fraction (RAF) and the mesomorphic phase contributions. As a consequence, in the following text and in Table 1, the amounts of the mesophase and the RAF will be jointly displayed as one entity so called $X_{ra}+X_{meso}$.

**Table 1.** TMDSC results: degree of crystallinity ($X_c$), degree of mobile amorphous phase (MAP) ($X_{ma}$), degrees of rigid amorphous fraction (RAF) and mesophase ($X_{ra} + X_{meso}$), mean temperature fluctuation related to the glass transition ($\delta T$), dynamic glass transition temperature ($T_\alpha$), cooperativity length at the dynamic glass transition temperature ($\xi_{T\alpha}$), fragility index (m), and volume contribution to fragility ($m - m_v$) with $\alpha_T/\kappa = 1.5$ MPa K$^{-1}$ for materials studied in this work

|     | $X_c$ (%) | $X_{ma}$ (%) | $X_{ra}+X_{meso}$ (%) | $\delta T$ (°C) | $T_\alpha$ (°C) | $\xi_{T\alpha}$ (nm) | m | m−m$_v$ |
|---|---|---|---|---|---|---|---|---|
| A | 0 ± 4 | 100 ± 2 | 0 ± 6 | 3.4 ± 0.2 | 57.5 ± 0.5 | 2.7 ± 0.3 | 155 ± 20 | 47 |
| UCW | 28 ± 4 | 66 ± 2 | 6 ± 6 | 4.6 ± 0.3 | 63.0 ± 0.5 | 2.0 ± 0.2 | 108 ± 10 | 19 |
| SEQ | 27 ± 4 | 70 ± 2 | 3 ± 6 | 4.1 ± 0.2 | 65.0 ± 0.5 | 2.3 ± 0.2 | 123 ± 15 | 29 |
| SB2 | 13 ± 4 | 86 ± 2 | 1 ± 6 | 2.6 ± 0.1 | 60.0 ± 0.5 | 3.1 ± 0.3 | 130 ± 15 | 69 |
| SB3 | 25 ± 4 | 66 ± 2 | 9 ± 6 | 4.8 ± 0.3 | 65.5 ± 0.5 | 2.0 ± 0.2 | 70 ± 10 | 19 |
| SB4 | 31 ± 4 | 62 ± 2 | 7 ± 6 | 5.1 ± 0.3 | 69.0 ± 0.5 | 1.8 ± 0.2 | 66 ± 10 | 14 |
| Th | 31 ± 4 | 59 ± 2 | 10 ± 6 | 4.7 ± 0.3 | 66.0 ± 0.5 | 1.8 ± 0.2 | 177 ± 20 | 14 |
| TCR | 31 ± 4 | 41 ± 2 | 28 ± 6 | 5.0 ± 0.3 | 63.0 ± 0.5 | 1.6 ± 0.2 | 174 ± 20 | 10 |

Being extensively studied in the literature for PLA and other semi-crystalline polyesters,[5,6,16,44] an isotropic thermally crystallized sample is chosen as a reference to understand microstructural modifications occurring in strain induced crystallized samples. In the spherulitic structure induced by thermal crystallization (*TCR*),[3] the degree of crystallinity and the content of RAF + mesomorphic phase are estimated around 30 %. This reveals a high coupling between the crystalline and the amorphous phases. Indeed it has been shown that when PLA is annealed at a low crystallization temperature (80°C), the probability is high for a simultaneous growth of RAF and crystalline content,[6,45] since the reorganization of polymeric segments in crystals is hindered by geometric restrictions due to the lack of chain mobility.[16] For the unidirectional drawing process *UCW* and the sequential unidirectional process *SEQ*, the phase contents are very similar, although the samples exhibit strongly different crystalline orientations. Indeed as previously seen on WAXS patterns from Figure 1, crystals are highly oriented along the draw direction for the UCW sample, while they are more randomized in the case of the SEQ sample, even if two preferential orientation directions, corresponding to the two draw directions, are observed.

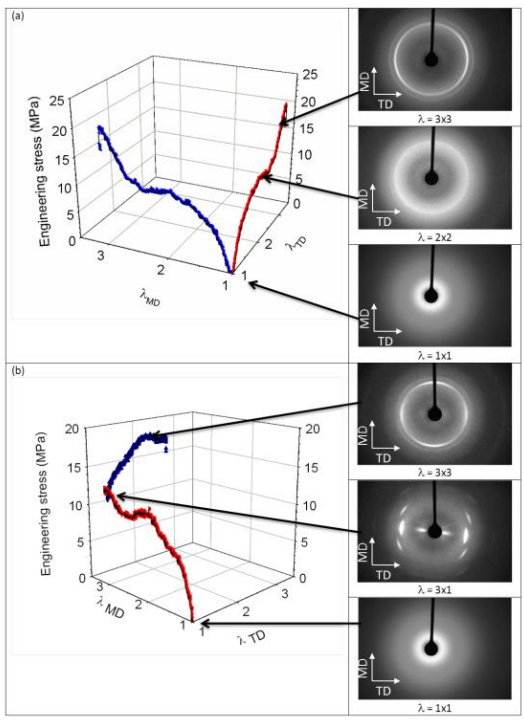

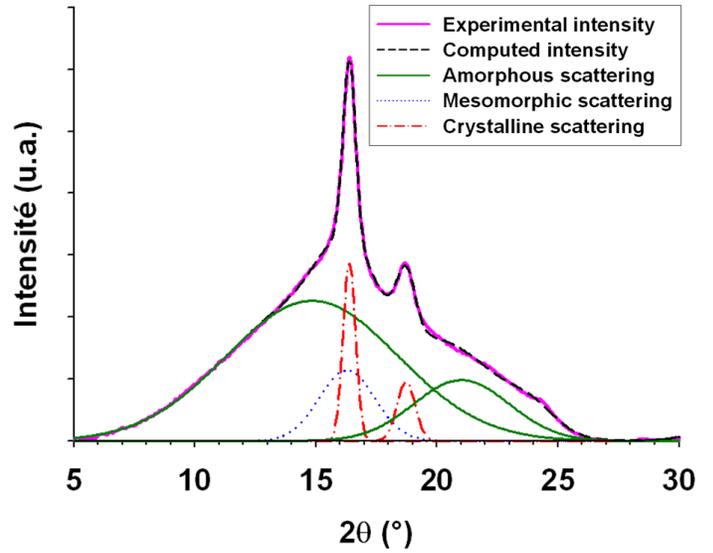

**Figure 1.** Mechanical behavior of PLA samples (a) simultaneously drawn (*SB*) and (b) sequentially drawn (*SEQ*) with the WAXS patterns obtained normal to the film plane at different drawing steps

**Figure 2.** Integrated intensity profiles of SB3

Like for thermal crystallization, their degree of crystallinity is close to 30%, which seems to be the maximum degree of crystallinity achievable for this grade of PLA. For simultaneous drawing processes, during *SB* drawing, the degree of crystallinity increases from 13% to 31% when the draw ratio increases from $\lambda = 2 \times 2$ to $\lambda = 4 \times 4$. One can consider that with the exception of the *SB2*, the maximum degree of crystallinity is reached in drawn materials. Like for *SEQ*, their RAF + mesophase content is weak and lower than 10%. Finally, no significant change concerning the phase content in *SB3* is observed before and after thermofixation (*Th*). However, this treatment influences the crystalline orientation in the material, as shown in Figure 3, where the azimuth intensity profiles of the (110)/(200) main crystalline reflection for the *SEQ*, *SB3*, and *Th* materials are compared. The microstructure of the SEQ material is highly oriented as revealed by the intensity maxima observed at $\varphi = 0°$ and $\varphi = 90°$, corresponding to the draw directions. The SB3 material exhibits the same profile, but the azimuthal intensity is significantly lower in the preferential directions indicating a more random distribution of the crystals. Worth noting is that WAXS analyses (results shown in supporting information) also reveal that in both cases crystals are orthotropically oriented—in other words, that strain-induced crystals preferentially lie in the film plane. Contrariwise, for the thermofixed material (Th), the azimuth profile is quasi-constant, revealing that the thermofixation process involves a randomization of the crystals' orientations.

**Thermal Analysis and Molecular Mobility**

The appearance of oriented crystallites combined to weak amount of RAF suggests changes in the interphase between crystals and MAP. Similar modifications may greatly modify the molecular mobility of the MAP as observed by several authors.[46,47] To investigate the effects of these modifications on the MAP, the dynamic glass transition was studied using TMDSC. The real part of the heat capacity and its derivative with temperature in the glass transition temperature vicinity are presented for some representative samples in Figure 4. First of all, a general increase of the glass transition temperatures can be noted, taking as reference the amorphous sample (*A*) with $Tg(A) = 58 °C$.

The weakest increase is observed for the early stage of the *SB* drawing process *Tg(SB2)* = 60°C; for the other samples with no discrimination between thermal or drawing crystallization process, the glass transition appears around 64°C ± 1°C; whereas for the most oriented sample *Tg(SB4)* = 69°C. $T_g$ values are mainly influenced by the crystalline degree. In order to study the molecular mobility of the MAP, we used the approach developed by Donth.[48] This approach has been widely used to analyze the molecular mobility in amorphous domains of complex systems at the dynamic glass transition.[25,31,49] Compared to other approaches, its main advantage is that the resulting fluctuation of the dynamic glass transition temperature can be obtained empirically by means of TMDSC. The cooperativity length $\xi_{T\alpha}$ is obtained according to the equation proposed by Donth:[48]

$$\xi_{T\alpha} = \left( \frac{\Delta(1/C_p)}{\rho(\delta T)^2} k_B T_\alpha^2 \right)^{1/3} \quad (5)$$

With δT as the mean temperature fluctuation related to the dynamic glass transition of one CRR, $k_B$ as the Boltzmann constant, ρ as the MAP density, and $C_p$ as the heat capacity at constant pressure. When applying the fluctuation-dissipation theorem (FDT) and modeling the out-of-phase component C"(T) or the derivative of C'(T), with temperature as a Gaussian (supposing temperature variable and frequency constant), δT corresponds to the standard deviation of this Gaussian,[50] and $T_\alpha$ corresponds to the maximum. In the case of semi-crystalline polymers, $\Delta(1/C_p)$ is estimated from the C' step normalized to the quantity of amorphous phase relaxing at the glass transition. The corresponding relaxation parameters are reported in Table 1.

The thermal crystallization at 80°C (*TCR*) on the amorphous film (*A*) induces a broadening of the glass transition, characterized by an increase of δT from 3.4 °C to 5.0 °C. The glass transition broadening is often correlated with a broadening of the relaxation time distribution of the MAP[51] and implies that the molecular dynamics of the amorphous phase have become more heterogeneous. This obviously influences the cooperativity which decreases from 2.7 nm for the *A* to 1.6 nm for the *TCR*. These differences are related to the microstructure of both materials. For the *TCR* material, as evidenced in previous papers by means of TMDSC[52], enthalpy of recovery study[53], and Polarized Optical Microscopy[52], the amorphous phase is totally trapped in spherulites. In this structure, the high coupling between phases generates a gradient of molecular dynamics depending on the macromolecule environment in the amorphous part. We assume that this change is related to the contribution of the geometrical restrictions of the amorphous segments—that is, the physical confinement of the MAP.[45,54,55,56]

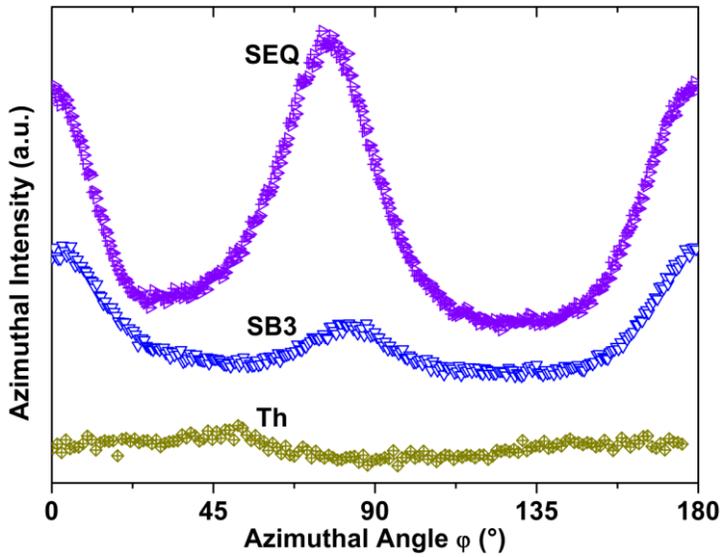

**Figure 3.** Azimuthal intensity profiles of the (200)/(110) diffraction ring for the *SB3*, *SEQ*, and *Th* samples

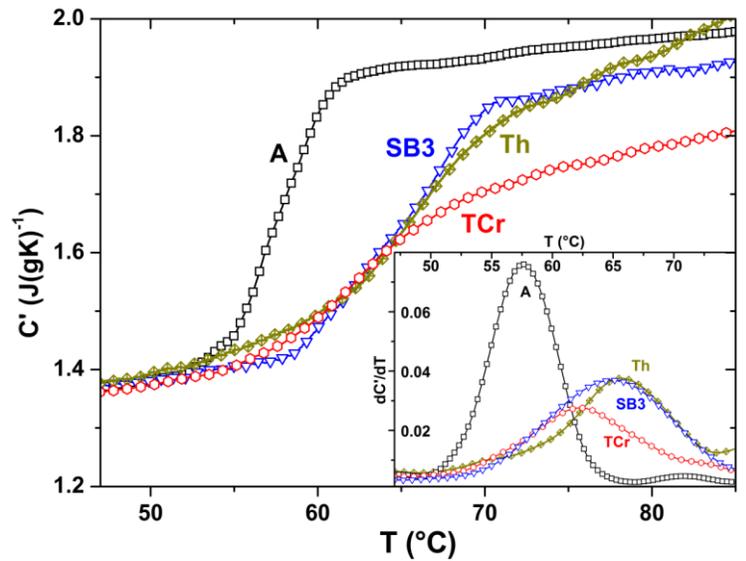

**Figure 4.** In-phase component (C') of the complex heat capacity and its derivative with temperature in inset (dC'/dT)

Similar results are obtained after *UCW* and *SEQ* drawings even if the cooperativity length is slightly higher ($\xi_{T\alpha}$ = 2.0 nm). These results are supported by previous ones obtained with drawn PET,[49,57,58,59] showing that drawing can also induce high heterogeneities in molecular dynamics. The decrease of the cooperativity length in drawn materials is mainly associated to two phenomena: the complexity of the environment that causes heterogeneities in the relaxation process so creates an effect close to confinement conditions, and the orientation of the macromolecules that causes the breaking of weak energy intermolecular bonds. . The relation between the intermolecular bond breaking and the drop of cooperativity is assumed by many authors due to their observations[60,61,62,63]. Regarding *SB* drawing, during the early stages of the process (*SB2*), no clear variation of $\xi_{T\alpha}$ can be observed. For SB3 and SB4, the structure becomes oriented and the degree of crystallinity increases; then $\xi_{T\alpha}$ decreases from 3.1 nm to 1.8 nm when the draw ratio increases from 2x2 to 4x4. For these draw ratios, the nature of the drawing process greatly influences the relaxation parameters. Finally, no drastic change is observed before (*SB3*) and after (*Th*) annealing of the 3x3 *SB* drawn sample. Small differences are imputed to the stabilization of the microstructure, with a slight increase of the degree of crystallinity. To complete the information given by the determination of the cooperativity length, the nature of the cooperative motions at the glass transition is investigated from the temperature dependence of the relaxation time, determined by means of BDS analysis. In Figure 5a, a 3-D plot of the imaginary ε″ part of the dielectric permittivity vs. frequency and temperature presents the wide frequency dependence of the relaxation behavior for sample *A*. In the low temperature range (-150° to 0°C), PLA exhibits a broad secondary relaxation process classically labeled β relaxation in the literature and assigned to twisting motions in the main chain.[64] Ren et al.[65] estimated the amplitude of such motions by an average twisting angle of around 11°. For temperatures ranging from 50°C to 100°C, the main α relaxation takes place, associated to the structural relaxation of PLA (dynamic glass transition). For temperatures higher than $T_g$ and at low frequencies, conduction effects appear, presenting a huge increase in ε". The inset of Figure 5a presents a topographic view of the α relaxation process, where the curvature of the temperature dependence of the relaxation rates taken at the maximum of the peak can be observed. In Figure 5b, the loss parts of the isochronal measurements (f = 1 kHz) for the same representative samples as in Figure 4 are given in the vicinity of the alpha relaxation.

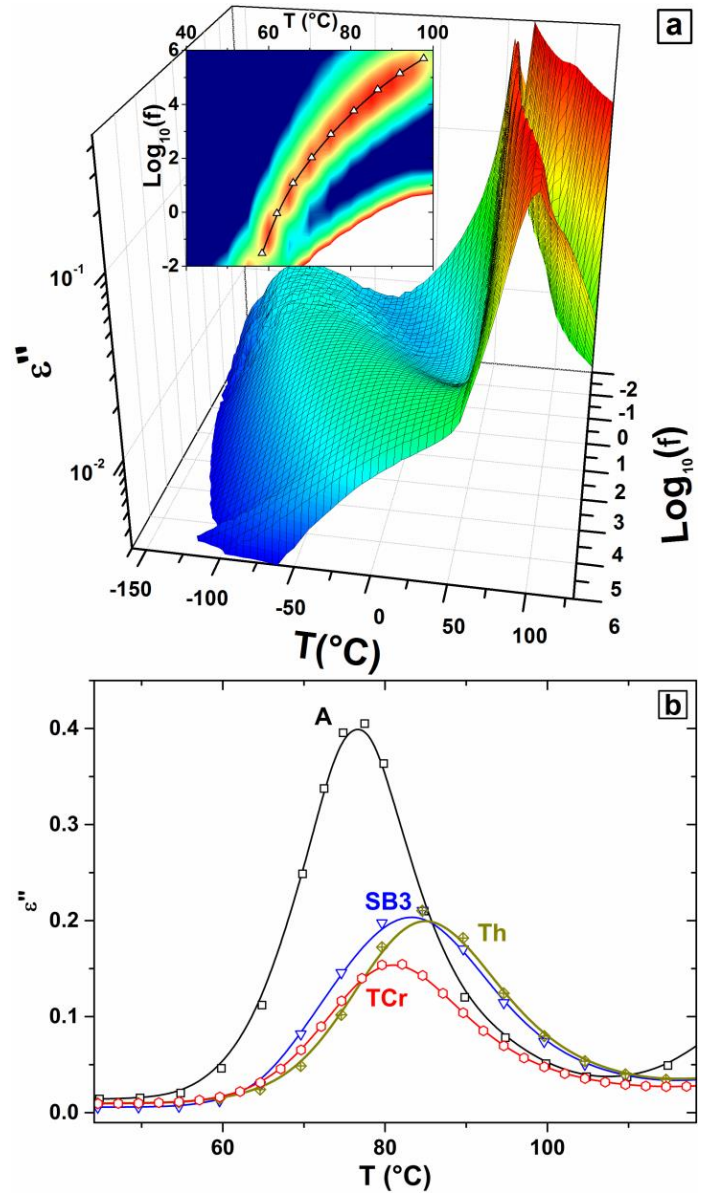

**Figure 5.** a) Example of BDS analysis for sample *A*: dielectric loss versus frequency and temperature; (inset) topographic view of the main relaxation temperature range of the α relaxation fitted with the VFT equation (full line). b) Isochronal measurements (f = 1 kHz) in the α relaxation temperature range.

This comparison exhibits the same trend as TMDSC: i) The amplitude of the loss peak decreases due to the MAP reduction; ii) The maximum of the peak shifts to higher temperatures in semi-crystalline samples; and iii) A broadening of the loss peak can be observed for semi-crystalline samples compared to the amorphous sample. The temperature dependence of the relaxation times for the α-relaxation of polymers is classically known to be curved versus 1/T, which one can describe by the Vogel-Fulcher-Tammann (VFT) equation.

$$\tau = \tau_0 \exp\left(\frac{DT_0}{T - T_0}\right) \quad (6)$$

Where D is a steepness parameter and $T_0$ is the reference Vogel temperature. Steepness is a valuable quantity to compare glassy dynamics of different systems. In the view of Angell et al.,[28] polymers are called "fragile" if the $\tau_\alpha(T)$ dependence deviates strongly from an Arrhenius-type behavior (equation 7) to fit VFT behavior, and "strong" if $\tau_\alpha(T)$ fits a linear dependence associated to an Arrhenius-type behavior:

$$\tau = \tau_0 \exp\left(\frac{E_a}{T}\right) \quad (7)$$

**Investigation of the Temperature Dependence of the Structural Relaxation**

The fragility index *m*, introduced by Angell et al.,[28] is an interesting parameter to quantify the deviation degree of the temperature dependence of the relaxation times from an Arrhenius behavior. This approach enables to compare glass-forming systems from polymers,[46,66] glass-forming liquids,[67] and metallic glasses.[68] This parameter can be determined using the temperature dependence of the relaxation time in a temperature-normalized scale:

$$m = \left.\frac{d\log_{10}(\tau)}{d(T/T_g)}\right|_{T=T_g} \quad (8)$$

As expected, relaxation rate ($f_{max}$) fits well into VFT law (eq. 6), and the broadening of the dissipation zone (highlighted in red in Figure 5a inset) can be observed as the relaxation rates increase. Figure 6 depicts for each sample the temperature dependence of the α-relaxation time in a temperature normalized scale in $T_g$ (i.e., Angell plot). The calculated fragility index *m* shows that initially "fragile" PLA, *m* = 155, reaches a "stronger" behavior during *SB* drawing as draw ratio increases (*m*= 130, 70, and 66 for *SB2*, *SB3*, and *SB4*, respectively). *UCW* and *SEQ* drawings lead also to a decrease in the fragility index, but the effect is less important for the *SEQ* drawing. This behavior is not identical in the annealed samples (*Th* and *TCR*) whether or not they have undergone a preliminary drawing stage. *Th* and *TCR* exhibit slightly higher values of fragility index than the amorphous material (respectively 177 and 174 versus 155). $m_V$ is the isochoric fragility, $\Delta V^\#$ is the activation volume (i.e., the excess volume needed for a single relaxation unit to relax), $\kappa$ is the compressibility, and $\alpha_T$ is the thermal expansion coefficient of the supercooled liquid at $T_g$. The ratio $\alpha_T/\kappa$ varies from 0.5 to 3.0 MPa / K for a wide range of glass formers,[74] and $\Delta V^\#$ is equal to about 4 % of the cooperativity volume. This leads to draw the limits of a theoretical value domain for the volume contribution $m - m_V$ as shown in Figure 7, where the fragility index is presented as a function of the cooperativity length for each studied material. Bras et al. even reported *m* values higher than 200 for low-temperature thermal cold-crystallization.[84] Many studies carried on other semi-crystalline polymers evidenced antagonist effects of the microstructure on the fragility. Cerveny et al.[85] have shown that the constrained amorphous phase in trans-1,4 polyisoprene exhibits higher fragility than the amorphous matrix (about 150 versus 115). On the other hand, crystallization at 100°C on initially amorphous PET generates a decrease of *m* from 133 to 65.[81] To explain these differences, Napolitano and Wubbenhorst[86] assumed that the effect of the amorphous phase confinement on the fragility index depends on the intrinsic chain flexibility: only polymers having low chain flexibility like PET exhibit a reduction of the segmental mobility. The influence of the chain orientation was studied in PVDF by Linares et al.[87] They reported very close fragility values between oriented and isotropic semi-crystalline polymers (respectively 60 and 64). Our results show, however, that the assumption of a direct correlation between the fragility index and the cooperativity length is hasty. Even correlation between the fragility and the crystalline phase ratio in PLA seems to be hazardous. Firstly, the fragility index can drastically vary between materials exhibiting the same degree of crystallinity (66 for *SB4* and 177 for *Th*). On the other hand, the fragility values can be close between an amorphous material and a material reaching its maximum degree of crystallinity (174 for *TCR* and 155 for *A*).

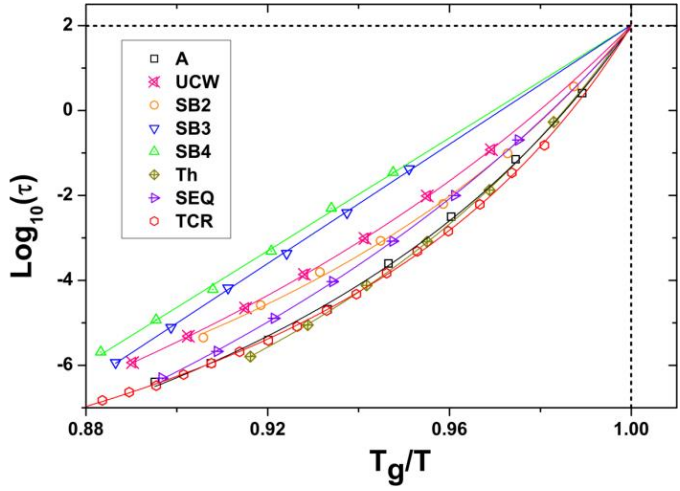

**Figure 6.** Values of $\text{Log}_{10}(\tau)$ versus $T_g/T$ for all samples: the full lines correspond to Arrhenius fits for 3x3 and 4x4 SB drawn PLA and to VFT fits for the other samples

In Figure 8, the evolution of cooperativity length and fragility index is reported as a function of the content of the crystalline phase ($X_c$), MAP ($X_{ma}$), and residual phases (i.e., the sum of mesophase and RAF [$X_{ra}$ + meso]). We already evoked that the high fragility values obtained for the *Th* and *TCR* materials break down the idea of a simple correlation between the degree of crystallinity and the fragility index. As previously discussed, it has been shown that depending on the polymer repeating unit or depending on the crystallization mode, an increase or decrease of the fragility index with an increase of the crystalline phase can be observed[81,85]. In our case also, no clear trend can be extracted from the fragility index variation with degree of crystallinity (Figure 8b). However, $\xi_{T\alpha}$ decreases as the degree of crystallinity increases (Figure 8a), which is supported by the literature.[88]

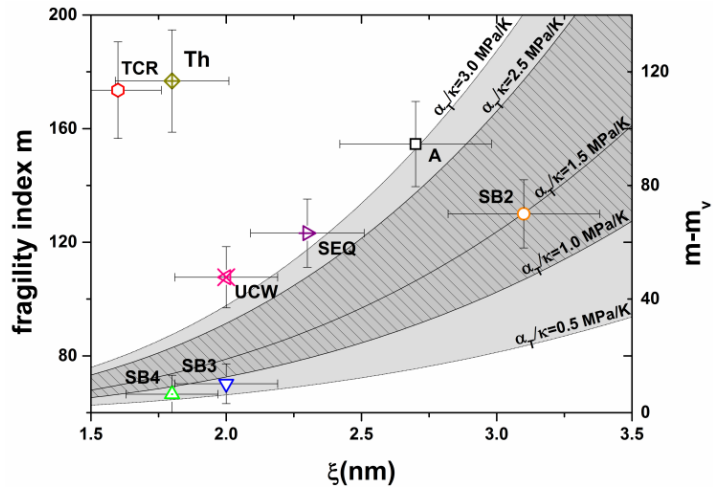

**Figure 7.** The colored dots are the experimental fragility index $m$ as a function of the cooperativity length $\xi$ for each material. The light grey area corresponds to the theoretical value domain of the "volume contribution to fragility" $m-m_V$ as a function of the cooperativity length, calculated according to equation 9, with 0.5 MPa/K < $\alpha_T/\kappa$ < 3.0 MPa/K, as defined by Hong et al. for a wide range of glass formers[74]. The hatched area is delimited by the values of 1.0 MPa/K < $\alpha_T/\kappa$ < 2.5 MPa/K and corresponds to the reduced domain of polymers that have been investigated by Hong et al[74].

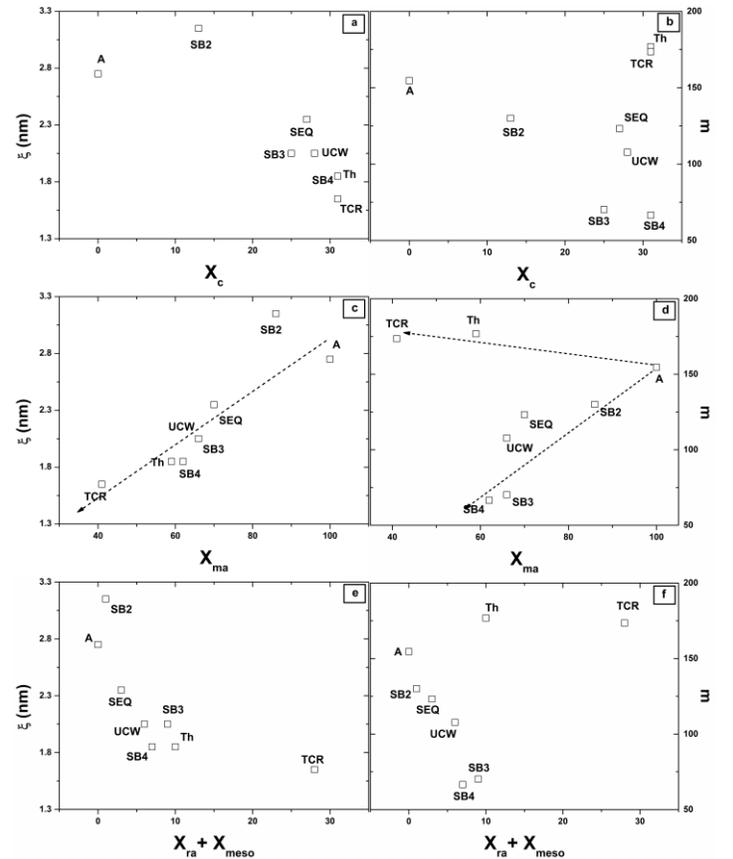

**Figure 8.** Cooperativity length $\xi$ versus (a) degree of crystallinity $X_c$, (c) degree of MAP $X_{ma}$ and (e) sum of degrees of RAF and mesophase ($X_{ra}$ + meso); fragility index $m$ versus (b) degree of crystallinity $X_c$, (d) degree of MAP $X_{ma}$ and (f) sum of degrees of RAF and mesophase ($X_{ra}$ + meso)

The reason mainly developed in these studies for the decrease in cooperativity length is the geometric restriction of the amorphous fraction between crystallites, inducing a geometrical confinement. Considering that the α relaxation process occurs in the MAP, one can suggest that the MAP content $X_{ma}$ is a more significant parameter to investigate a global-size effect on the relaxation parameters.[49] Correlation with the cooperativity length is direct: the greater the MAP, the higher the CRR size (Figure 8c). On the contrary, fragility values are split into two tendencies. The fragility of the annealed materials increases with the reduction of the MAP, while it decreases for drawn samples (Figure 8d). In the literature, $m_V$ is defined as a pure energetic parameter and depends on the nature of the inter- and intra-molecular bonds and on intra-molecular degrees of freedom.[74] It is also clearly associated to the packing efficiency of the amorphous chain.[77] Regarding *SB3*, the previously observed crystalline texturation leads to reasonably picture an orientation of the amorphous part. The packing efficiency of this oriented amorphous fraction increases with the draw ratio, causing the general decrease of the fragility index and more specifically $m_V$. When the same sample undergoes a thermofixation (*Th*), the WAXS analysis (Figure 3) highlights the disappearance of the crystalline texture. Concerning the amorphous fraction, we may reasonably propose that the initial induced orientation is widely suppressed, leading to a more isotropic glassy structure. This relaxation of the amorphous phase leads to a rearrangement of macromolecules conformation into a random coil, with a packing efficiency much lower than the amorphous fraction of *SB3* (Scheme 1). As supported by the work of Ou et al.[90] the postdrawing annealing mainly produces a reorientation of crystallites without change in their average repeat distance (Scheme 1) and thus in the geometrical confinement. This result is in agreement with the stability of the cooperativity length before and after thermofixation. The results obtained for the *TCR* material are related to the specific nature of the intra-spherulitic MAP. This phase is separated from the crystals by the RAF acting as a buffer zone[64]. The remarkable increase in the degrees of freedom of the relaxing part of the macromolecule is a secondary effect of the local restriction of the mobility occurring in the RAF. Thus the molecular mobility in the intra-spherulitic amorphous phase can be pictured by considering segments subjected to geometrical limitations in the RAF and others exhibiting low packing efficiency in the MAP. For TCR also, the cooperativity length and so the $m—m_V$ remain low due to the geometrical confinement of the intra-spherulitic MAP.

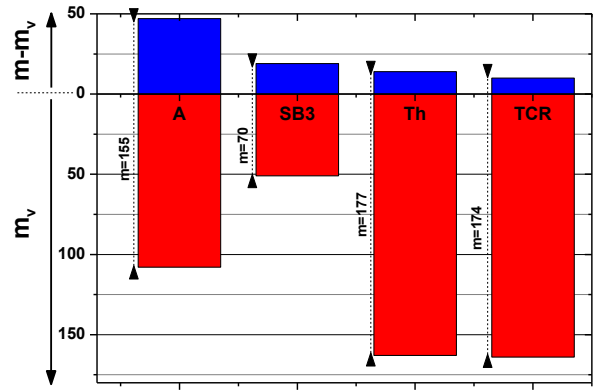

**Figure 9.** Repartition of the volume ($m—m_V$) and thermal ($m_V$) contributions to the fragility index $m$

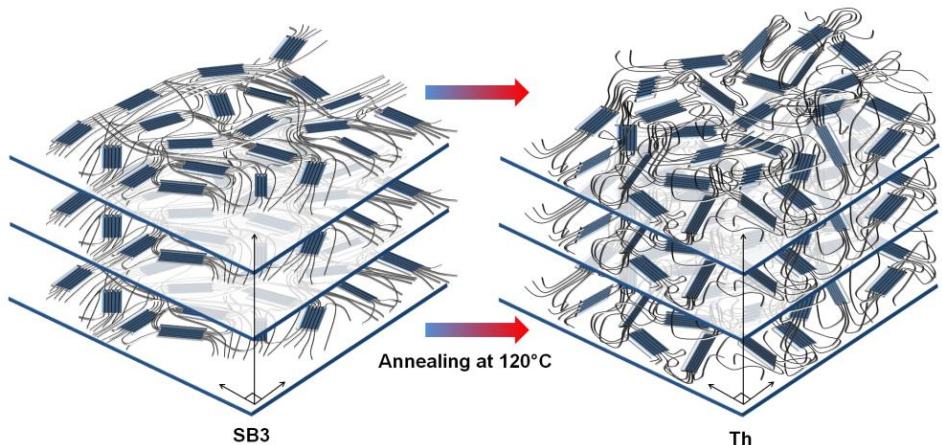

**Scheme 1.** Schematic view of the effect of thermofixation on the SB3 sample microstructure.

# CONCLUSIONS

The macromolecular orientation generated by drawing processes above $T_g$ on amorphous polylactide leads to the appearance of oriented crystals and to geometric restrictions in the amorphous phase that limit the segmental motions. This leads to a global decrease of the cooperativity length which is similar to a pure geometrical confinement effect and may be attributed to the prevalence of covalent bonds compared to weak physical bonds in such structures. This also changes the temperature dependence of the relaxation times, which evolves from an original VFT behavior towards an Arrhenius behavior. The resulting decrease of the fragility index is attributed to the increase of the amorphous chain packing efficiency in the glassy state. During a subsequent thermofixation, the initial constraints of the amorphous phase are widely slackened. No effect on the cooperativity length is reported, but a huge increase of the fragility index related to the molecular reorganization into a random coil is observed. These relaxation parameters are identical to those of intra-spherulitic MAP, which is decoupled from the crystalline phase by the "buffer effect" of the RAF and exhibits high molecular mobility. This study proves the non-existence of an absolute correlation between fragility and cooperativity length in semi-crystalline polymers. It also gives credit to the assumption that the fragility index is the sum of a volume contribution which is directly related to cooperativity, and an energetic contribution that depends on the degrees of freedom in the relaxing part of the amorphous phase. The existence of distinct molecular dynamics for similar microstructures is a step forward to the understanding of singular properties related to the behavior of the polymer amorphous phase.

ACKNOWLEDGMENT

The authors would like to thank the region Haute Normandie for their financial support and the acquisition of the Broadband Dielectric Spectrometer and Mr Araujo Steven and Mr Nguyen Van Hung for their participation to the experimental part of the work. Financial support from Région Nord Pas de Calais and European FEDER for SAXS-WAXS laboratory equipment is also gratefully acknowledged.